\documentstyle[epsfig,preprint,aps]{revtex}
\newcommand{\gsi}        {$\rm^{1}$}
\newcommand{\warsaw}     {$\rm^{2}$}
\newcommand{\clt}        {$\rm^{3}$}
\newcommand{\heidelberg} {$\rm^{4}$}
\newcommand{\bucarest}   {$\rm^{5}$}
\newcommand{\zagreb}     {$\rm^{6}$}
\newcommand{\itep}       {$\rm^{7}$}
\newcommand{\budapest}   {$\rm^{8}$}
\newcommand{\korea}      {$\rm^{9}$}
\newcommand{\dresde}     {$\rm^{10}$}
\newcommand{\kur}        {$\rm^{11}$}
\newcommand{\ires}       {$\rm^{12}$}

\begin{document}
\draft

\title{ 
Direct comparison 
of phase-space distributions of K$^{-}\, $ and K$^{+}\, $ mesons  
in heavy-ion collisions at SIS energies -
evidence for in-medium modifications of kaons ? \\
} 

\author{K.~Wi\'{s}niewski\gsi$^{\rm ,}$\warsaw,
P.~Crochet\clt,
N.~Herrmann\gsi$^{\rm ,}$\heidelberg,
A.\,Andronic\gsi,
R.~Averbeck\gsi,
A.~Devismes\gsi,
C.~Finck\gsi,
A.~Gobbi\gsi,
O.~Hartmann\gsi,
K.D.~Hildenbrand\gsi, 
P.~Koczon\gsi,
T.~Kress\gsi,
R.~Kutsche\gsi,
Y.~Leifels\gsi$^{\rm ,}$\heidelberg,
W.~Reisdorf\gsi,
D.~Sch\"ull\gsi, 
J.P.\,Alard\clt,
V.\,Barret\clt,
Z.\,Basrak\zagreb,
N.\,Bastid\clt,
I.\,Belyaev\itep,
A.\,Bendarag\clt,
G.\,Berek\budapest,
R.\,\v{C}aplar\zagreb,
N.\,Cindro\zagreb,
P.\,Dupieux\clt,
M.\,D\v{z}elalija\zagreb,
M.\,Eskef\heidelberg,
Z.\,Fodor\budapest,
%
Y.\,Grishkin\itep,
B.\,Hong\korea,
J.\,Kecskemeti\budapest,
Y.J.\,Kim\korea,
M.\,Kirejczyk\warsaw,
M.\,Korolija\zagreb,
R.\,Kotte\dresde,
M.\,Kowalczyk\warsaw,
A.\,Lebedev\itep,
K.S.\,Lee\korea,              
V.\,Manko\kur,
H.\,Merlitz\heidelberg,
S.\,Mohren\heidelberg,
D.\,Moisa\bucarest,
W.\,Neubert\dresde,
A.\,Nianine\kur,
D.\,Pelte\heidelberg,
M.\,Petrovici\bucarest,
C.\,Plettner\dresde,
F.\,Rami\ires,
B.\,de Schauenburg\ires,
Z.\,Seres\budapest,
B.\,Sikora\warsaw,
K.S.\,Sim\korea,
V.\,Simion\bucarest,
K.\,Siwek-Wilczy\'nska\warsaw,
V.\,Smolyankin\itep,
A.\,Somov\itep,
M.\,Stockmeier\heidelberg,
G.\,Stoicea\bucarest,
M.\,Vasiliev\kur,
P.\,Wagner\ires,
D.\,Wohlfarth\dresde,
J.T.\,Yang\korea,
I.\,Yushmanov\kur,
A.\,Zhilin\itep\\
the FOPI Collaboration 
}

\address{
\gsi~Gesellschaft f\"ur Schwerionenforschung, Darmstadt, Germany\\
\warsaw~Institute of Experimental Physics, Warsaw University, Poland\\
\clt~Laboratoire de Physique Corpusculaire, IN2P3/CNRS
and Universit\'{e} Blaise Pascal, Clermont-Ferrand, France\\
\heidelberg~Physikalisches Institut der Universit\"at Heidelberg, Heidelberg, Germany\\
\bucarest~National Institute for Nuclear Physics and Engineering, Bucharest, Romania\\
\zagreb~Rudjer Boskovic Institute, Zagreb, Croatia\\ 
\itep~Institute for Theoretical and Experimental Physics, Moscow, Russia\\
\budapest~KFKI Research Institute for Particle and Nuclear Physics, 
Budapest, Hungary\\
\korea~Korea University, Seoul, South Korea\\
\dresde~Forschungszentrum Rossendorf, Dresden, Germany\\
\kur~Kurchatov Institute, Moscow, Russia \\
\ires~Institut de Recherches Subatomiques, IN2P3-CNRS and Universit\'e
Louis Pasteur, Strasbourg, France \\
}

\maketitle

\begin{abstract}
The ratio of K$^{-}\, $ to K$^{+}\, $ meson yields has been measured in the 
systems
$^{96}$Ru+$^{96}$Ru at 1.69 A GeV, 
$^{96}$Ru+$^{96}$Zr at 1.69 A GeV, and 
$^{58}$Ni+$^{58}$Ni at 1.93 A GeV incident beam kinetic energy.  
The yield ratio is observed to vary across the measured phase space.  
Relativistic 
transport-model calculations indicate that the data are best understood if 
in-medium modifications of the kaons are taken into account.
\end{abstract}

\vspace{1.cm}


Recently there has been considerable effort, both experimentally and 
theoretically, to investigate changes of hadron properties in
a hot and dense nuclear medium.  In particular, a variety of
theoretical approaches consistently  predict that the effective 
mass of kaons increases slightly with increasing baryon density, while the 
mass of antikaons is expected to drop substantially~\cite{sch97}. 
This phenomenon could lead to the formation of a kaon condensate in a dense 
hadronic environment~\cite{nel87}, which in turn would effect the nuclear 
equation of state, and have consequences for the physics of neutron stars 
\cite{bro94}.  The modifications of the properties of kaons in a 
hadronic medium 
might originate from the partial restoration of the chiral symmetry of QCD 
\cite{lut94}.

The question whether the kaon masses are modified in a dense hadronic 
environment can be addressed experimentally with studies on kaons produced 
in heavy-ion collisions at bombarding energies around 1-2 GeV per 
nucleon, which is close to the production threshold in elementary, 
nucleon-nucleon reactions (1.6 and 2.5 GeV for K$^{+}$ and 
K$^{-}$ mesons, respectively).  At these beam energies, kaons are most 
likely  produced in the early stage and in the central region of the 
collision~\cite{fan94}, where densities of up to 3 times the normal nuclear 
matter density and temperatures in the order of 100 MeV can be reached 
\cite{sto86}. The production rate can be influenced not only by the  
surrounding medium (e.g., its density) or the properties 
of the nucleons and their resonances, but also by possible changes of the 
kaon properties themselves.

The observation of the enhanced K$^{-}$ meson yield at midrapidity in 
heavy-ion collisions with respect to the elementary, nucleon-nucleon reactions 
\cite{bar97} is a very interesting signature, possibly related to a 
substantial in-medium drop of the effective mass of K$^{-}$ mesons. 
A change of the effective mass of a particle 
can be understood as an effect 
of a density dependent potential.  
Gradients of this potential cause forces that act on particles.
While antikaons are attracted into regions of high baryon 
density, kaons are repelled from these regions.  This effect offers an 
explanation~\cite{li95}  for the characteristic directed side-flow patterns 
of K$^{+}$ mesons that are observed experimentally~\cite{rit95}. 
It also predicts characteristic changes of the final-state phase-space 
distributions of kaons and antikaons~\cite{li97}. 

In this Note we report on measurements of K$^{+}$ and K$^{-}$ mesons 
produced in $^{96}$Ru+$^{96}$Ru/$^{96}$Zr collisions at 1.69 A GeV and 
in $^{58}$Ni+$^{58}$Ni collisions at 1.93 A GeV 
incident beam kinetic energy.  The experiments were performed at SIS/GSI, 
using the FOPI experimental setup~\cite{fop95} which 
allows for a simultaneous measurement of all charged 
reaction products. 
Thus final-state distributions of the  
particles can be directly compared within the same event sample 
and with the same acceptance.
Results on the phase-space population of $\pi^\pm$, $p$, $d$, and K$^{+}$
in the Ni+Ni experiment were reported elsewhere~\cite{bes97,hon98}.
Here, we show, for the first time in this energy regime,
the ratio of K$^{-}$ to K$^{+}$ meson yields in the backward hemisphere.  
We extract the ratio across a 
relatively wide region of phase 
space, which provides high sensitivity to 
the dynamics of the propagation of kaons through the medium.
We observe that the phase-space distributions of K$^{-}$ and K$^{+}$ 
mesons differ and discuss the origin of this effect.

An ensemble of events biased to small impact parameters has been selected by 
requiring high charged-particle multiplicity in the polar-angle 
range $7^\circ < \Theta_{Lab} < 30^\circ$ on the trigger
level.  For the Ni+Ni experiment, 
4.7$\cdot$10$^6$ events were selected, corresponding to the centralmost  
11\% of the total geometrical cross section.  
Since no difference in strangeness production was found 
between the Ru+Ru and Ru+Zr systems~\cite{WIS99}, 
the accumulated statistics 
was combined for a total of 7.7$\cdot$10$^6$ events, 
corresponding to the centralmost 14\% of the geometrical cross section.

The identification of K$^{+}$ and K$^{-}$ mesons with the FOPI detector 
relies upon the information on specific energy loss and track 
curvature in the Central Drift Chamber (CDC), and on the measurements 
of Time-of-Flight (ToF) in the surrounding Plastic Scintillator Barrel 
detector~\cite{fop95}.  
The acceptance is thus restricted to the polar-angle range in 
the laboratory reference frame 39$^\circ < \Theta_{lab} < $130$^\circ$ and 
to transverse momenta $p_t >$\,0.1\,GeV/c.  The finite detector resolution 
limits the identification of K$^{+}$ mesons to laboratory momenta 
$p_{lab}<0.5$\,GeV/c. Due to the much lower K$^{-}$ yield compared 
to that of K$^{+}$ meson, the background contamination in the former 
increases more rapidly with momentum. In order to eliminate possible 
distortions due to a momentum dependent background contribution, the 
considered momentum range for K$^{-}$ meson identification is restricted 
to $p_{lab}$ below 0.32 and 0.34\,GeV/c in case of the Ru+Ru/Zr and Ni+Ni 
experiments, respectively.  
At $p_{lab}$ $\simeq$ 0.32 (0.34)\,GeV/c the measured K$^{-}$ 
meson yields have less than 20\% background contamination and the K$^{+}$ 
meson yields are practically background free (the contamination is less 
than 5\%).  

\begin{figure}[htb]
 \begin{center}
  \epsfig{file=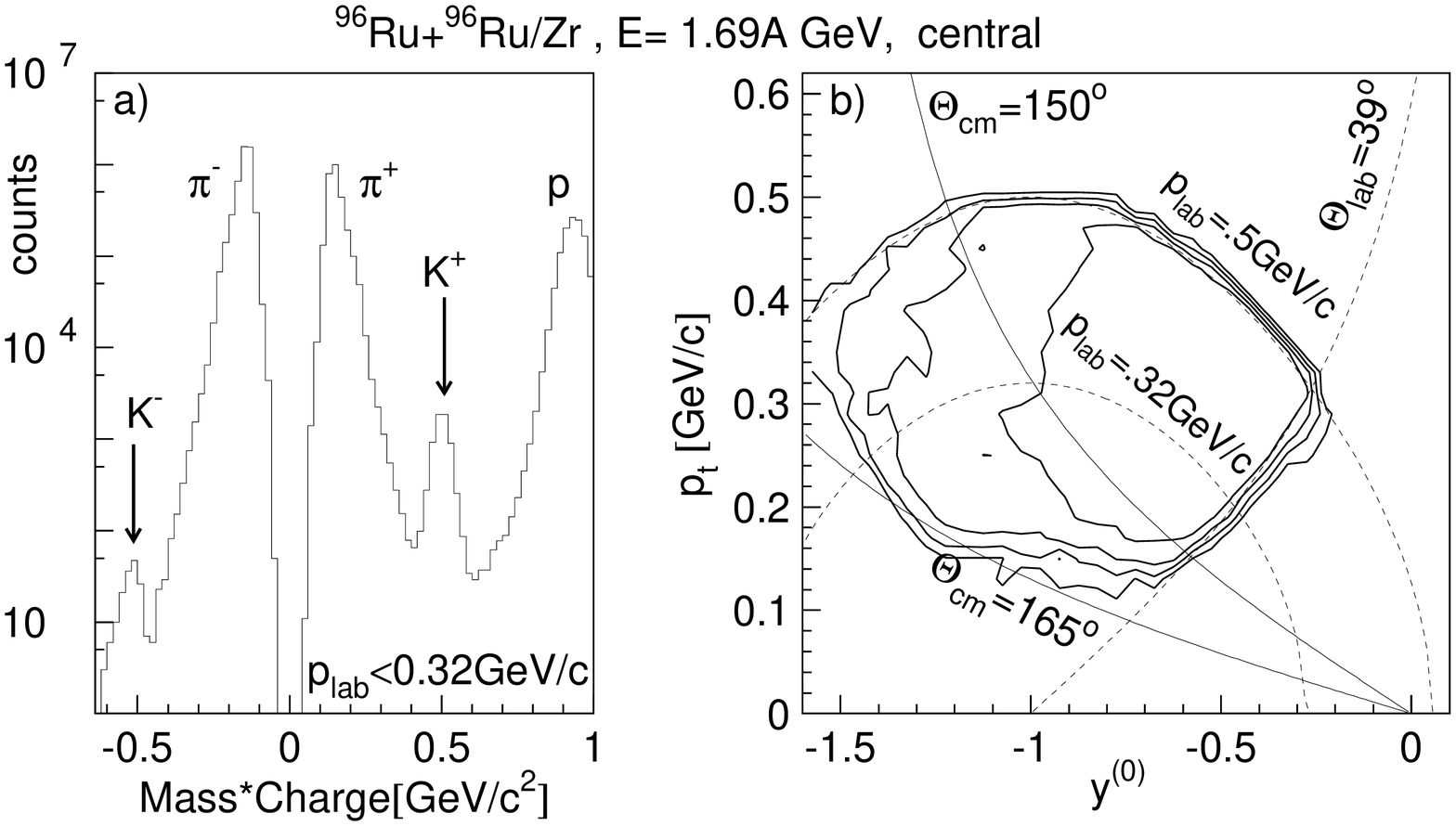,width=1.\linewidth}
 \end{center}
 \caption{
   (a) Mass spectrum of charge $\pm$1 particles measured in the Ru+Ru/Zr
   experiment.
   (b) K$^{+}$ meson phase-space distribution within the acceptance of the
   FOPI detector. The meaning of the lines is explained in the text.}
 \label{mass-acc} 
\end{figure}

With the different upper momentum limits for positive and negative 
kaons, mentioned above, 
around 26000 K$^{+}$ and 240 K$^{-}$ mesons have been identified in the Ni+Ni 
experiment.  The combined statistics of the Ru+Ru/Zr experiment is about 40000 
K$^{+}$ and 220 K$^{-}$ mesons.  The mass spectrum of particles with 
charge $\pm$1 measured in the Ru+Ru/Zr experiment is shown in 
Fig.\ref{mass-acc}.a.  
Peaks from K$^{+}$ and K$^{-}$ mesons are clearly visible.  
The portion of the phase space populated by K$^{+}$ mesons 
measured in the Ru+Ru/Zr experiment is shown in Fig.\ref{mass-acc}.b 
in terms of normalized rapidity 
($y^{(0)}=y^{lab}/y^{CM}-1$, where $y^{CM}$ is half of the beam rapidity) 
and transverse momentum. 
In this representation -1 and 0 on the rapidity axis correspond to the target 
and the 
midrapidity, respectively.  The yield of K$^{+}$ mesons is depicted by the 
contour lines on a linear scale.  The geometrical limit at 
$\Theta_{lab}$=39$^\circ$ and the upper $p_{lab}$ limits for K$^{+}$ and 
K$^{-}$ meson identification are depicted by dashed lines.  
The solid lines show the polar-angle range in the center-of-mass (c.m.) 
reference frame 
150$^\circ < \Theta_{cm} <$165$^\circ$, which will be referred to later.

To quantitatively examine and compare the phase-space distributions of 
K$^{+}$ and K$^{-}$ mesons, where for the latter case low statistics 
does not allow {to extrapolate the measured yields to 
experimentally not accessible regions of the phase space}, 
we study the {\em ratio} of K$^{-}$ to K$^{+}$ meson yields 
in the limited phase-space region defined by the K$^{-}$ meson 
identification. 
This offers two advantages  with 
respect to analysing the single particle distributions. 
(i) Experimental difficulties, 
like detection efficiencies and acceptance deficiencies, cancel to a large 
extent~\cite{WIS99}. 
(ii) In-medium effects act in opposite ways on K$^{-}$ and K$^{+}$ 
mesons, hence the ratio should reveal these more clearly.  
                 
\begin{figure}[htb]
 \begin{center}
  \epsfig{file=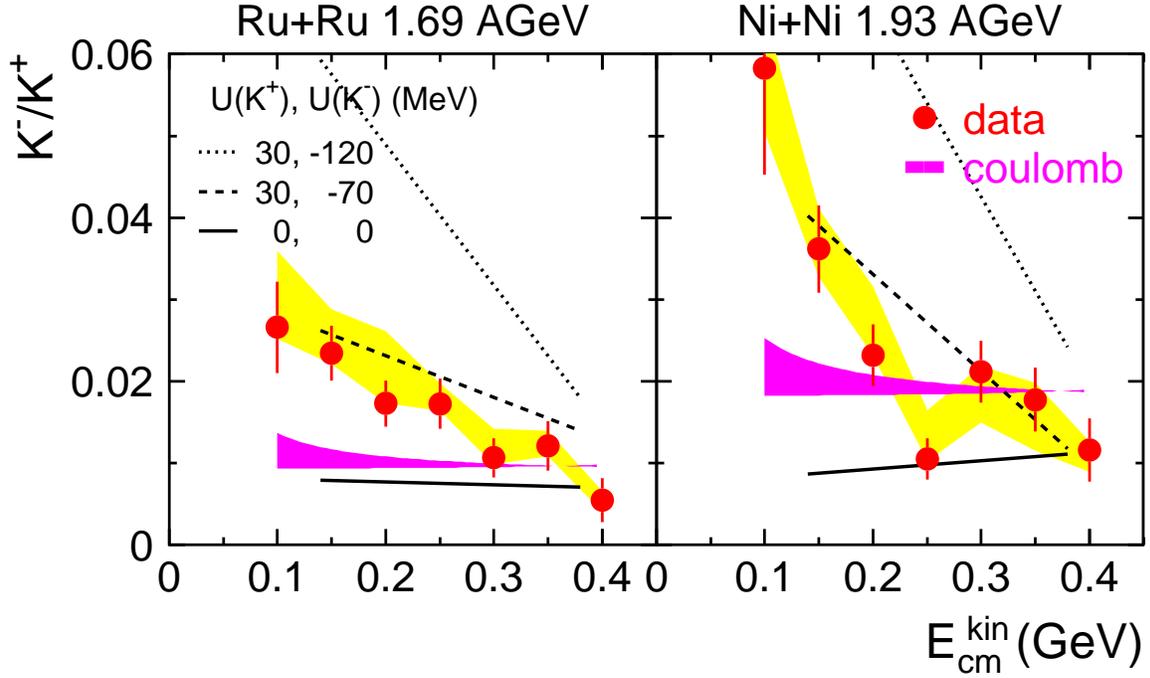,width=1.\linewidth}
 \end{center}
 \caption{The K$^{-}$/K$^{+}$ ratio as a function of $E^{kin}_{cm}$ in the
   Ru+Ru/Zr (left) and Ni+Ni experiments (right). 
   The data are extracted in the polar-angle range 
   150$^\circ < \Theta_{cm} <$165$^\circ$. 
   The light-grey shaded areas {correspond to the estimate of 
   systematic errors.} 
   {The lines depict predictions of the RBUU transport model 
   with different strength ${\rm U}(\rho=\rho_\circ)$ of the in-medium 
   (anti)kaon potentials at normal nuclear matter density.} 
   Statistical uncertainties of the predictions are similar to 
   those of the experimental data. 
   The horizontal dark-grey shaded areas show the results of numerical 
   simulations carried out in order to estimate the
   influence of the Coulomb potential on the K$^{-}$/K$^{+}$ ratio.}
 \label{ecm}
\end{figure}

\begin{figure}[htb]
\begin{center}
\epsfig{file=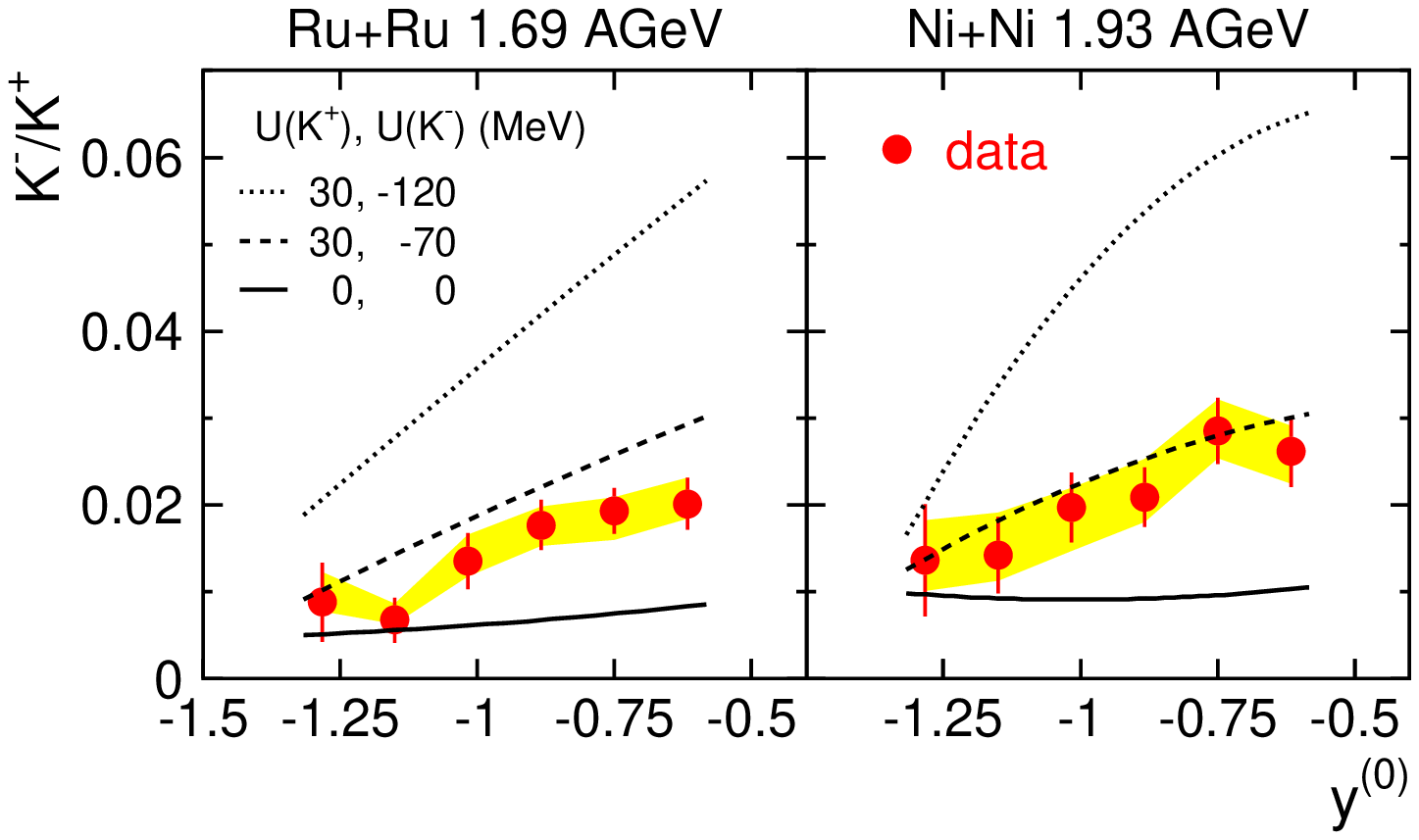,width=1.\linewidth}
\end{center}
 \caption{The K$^{-}$/K$^{+}$ ratio as a function of $y^{(0)}$
   in the Ru+Ru/Zr (left) and Ni+Ni (right)
   experiments. 
   The light-grey shaded areas {correspond to the estimate of
   systematic errors.}
   The lines depict predictions of the RBUU transport model 
   with different strength ${\rm U}(\rho=\rho_\circ)$ of the in-medium 
   (anti)kaon potentials at normal nuclear matter density.
   The results are filtered through the geometrical acceptance 
   of the detector. }
 \label{y0}
\end{figure}

In FOPI, the efficiency for 
particle detection is given by the tracking efficiency in the CDC and the 
matching efficiency with the ToF Barrel. 
Possible systematic bias on the measured K$^{-}$/K$^{+}$ ratio 
was estimated using a Monte-Carlo simulation in which the full
detector response was modelled with the GEANT package~\cite{brun78}.
The simulated data were analysed in the same way as the experimental data.
Comparing the output of the simulation to its input, the final 
K$^{-}$/K$^{+}$ ratio was found to be overestimated by 15\%, independently 
of transverse momentum and rapidity.  
This effect is attributed to different efficiencies of the track finding 
for positively and negatively charged particles due to the geometry of the 
CDC. A similar asymmetry of the efficiency 
was reported in~\cite{pelte97fig} for $\pi^+$ and $\pi^-$ mesons. 
All data points shown in the following figures are corrected for this
systematic bias, i.e., all ratios are reduced to 87\% of the directly 
measured value. Furthermore, the systematic uncertainties due to 
the identification criteria ($<$30\%) and the background contamination 
($<$20\%) were estimated by varying the conditions that were imposed on track 
parameters in order to select K$^{+}$ and K$^{-}$ mesons. 
These systematic errors are depicted by light-grey shaded areas 
in Fig.\ref{ecm} and \ref{y0}. 
The systematic distortion due to the nuclear scattering of K$^{-}$ mesons
in the target material is neglected, since for the targets used 
in the experients (1\% interaction length in the beam direction 
and a transverse diameter of about 12 mm), 
this interaction probability is estimated to be below 3\%. 

Since produced particles, and especially kaons, are found to be emitted 
almost isotropically in the c.m. system~\cite{bes97}, we plot 
in Fig.\ref{ecm} the K$^{-}$/K$^{+}$ ratio as a function of 
the kinetic energy in the c.m. reference frame ($E^{kin}_{cm}$) 
for both the Ru+Ru/Zr (a) and the Ni+Ni (b) experiments.  The polar-angle 
range 150$^\circ < \Theta_{cm} <$165$^\circ$ has been chosen since there the 
kinetic-energy acceptance is largest (see Fig.\ref{mass-acc}.b).  
We observe that the ratio rises towards small $E^{kin}_{cm}$.  

In order to test whether the effect can be caused by the electric 
field of the hadronic fireball formed in the collision, we performed 
numerical calculations following the studies on the influence of the 
Coulomb potential on charged-particle spectra reported in~\cite{ala97}. 
We followed the propagation of K$^{-}$ and K$^{+}$ mesons in an 
electric field of a net charge $Z$, which was originally homogeneously 
distributed in a sphere of a $R_f$ radius, and expanded radially with a 
$\beta_{rad}$ flow velocity.  Initially, the K$^{-}$ and K$^{+}$ mesons were 
evenly distributed within the expanding volume, and had identical energy 
spectra that corresponded to an isotropic emission from a midrapidity source of 
a $T_f$ temperature.  The yields of K$^{-}$ and K$^{+}$ mesons were 
arbitrarily normalized in order to allow a direct comparison with the experimental 
data. 
The parameters of the simulation (the total 
charge $Z$, the expansion velocity $\beta_{rad}$, the temperature 
$T_f$, and the radius $R_f$) were varied in a reasonable range 
in order to model possible freeze-out conditions.
The dark-grey shaded areas in Fig.\ref{ecm} 
correspond to the results obtained with different sets of parameters. 
Comparing this  to the data, we conclude that the influence of the Coulomb 
potential of the net positive charge of colliding ions on the K$^{-}$/K$^{+}$ 
ratio is too small to account for the observed relative narrowing (widening) of the 
K$^{-}$ (K$^{+}$) meson final-state phase-space distributions.

In Fig.\ref{y0} we plot the K$^{-}$/K$^{+}$ ratio as a function of 
$y^{(0)}$, i.e., in the direction parallel to the beam axis. 
The result is biased by the detector acceptance, but has the best statistical 
significance, which is given by the error bars attached to the data points.
We observe that the K$^{-}$/K$^{+}$ ratio rises towards midrapidity.
How representative the trend in the data is for the unbiased rapidity 
density distribution 
depends on the variation of the ratio as function of transverse 
momentum for fixed rapidity. For a meaningful comparison with the model 
predictions (see below), an acceptance filter was applied to the results 
of the model, taking into account the angular boundaries and 
the maximum laboratory momentum. 
The distortion due to acceptance turned out to be smaller than 20\% 
in the case of the rapidity dependence of the ratio shown in Fig.\ref{y0}. 
The kinetic-energy dependence of the ratio presented in 
Fig.\ref{ecm} is not biased by the acceptance in the considered 
range of $E^{kin}_{cm}$. 

Since the elementary reaction threshold is higher for K$^{-}$ than it is for
K$^{+}$ mesons (2.5 and 1.6 GeV respectively), 
one might expect that at the time of production, the 
average kinetic energies are lower for K$^{-}$ than for 
K$^{+}$ mesons. 
However, when the incident beam energy is far below threshold, 
the production of K$^{-}$ is dominated by channels involving 
intermediate baryon resonances and/or pions.  
According to Relativistic Boltzmann-Uehling-Uhlenbeck (RBUU) 
transport-model calculations~\cite{li98,cas97}
owing to this mechanism the initial 
momentum-space distribution of K$^{-}$ is even wider than that 
of K$^{+}$ mesons. In addition, particles are  
further rescattered, which tends to equalize   
their phase-space distributions. 
The solid lines in Fig.\ref{ecm} and \ref{y0} show the values of the K$^{-}$/K$^{+}$ 
ratio predicted by the RBUU transport model~\cite{cas97} when describing the 
kaon scattering in a fashion consistent with the 
free particle properties. No significant dependence of 
the ratio on $y^{(0)}$ and $E^{kin}_{cm}$ is found, 
which results in a manifest contradiction to the data.
The same conclusion has been drawn with another realization of 
a RBUU-type model~\cite{li98}.

Finally, we try to explain the variation of the K$^{-}$/K$^{+}$ ratio 
in the phase space by in-medium modifications of kaon properties.  
Dashed and dotted lines in Fig.\ref{ecm} and \ref{y0} show the values of the ratio 
predicted by the RBUU model when in-medium effects are taken into account
by a linear dependence of the in-medium potential on density~\cite{cas97}. 
Two scenarios with different strength ${\rm U}(\rho=\rho_\circ)$ 
of the in-medium kaon potentials at normal nuclear matter density 
are presented. 

The attractive K$^{-}$ potential influences the results 
shown in Fig.\ref{ecm} and \ref{y0} in a systematic fashion:
with increasing depth of the K$^{-}$ potential, the K$^{-}$/K$^{+}$ ratio 
increases on average, and in addition the slope  of the ratio with 
respect to rapidity and kinetic energy rises. 
While the first effect is caused by the corresponding
drop of the effective mass, the rapidity and kinetic-energy dependencies 
are generated by the forces originating from the gradients of the potentials. 
Similar observations have been made on the $p_t$ 
dependence of the K$^{-}$/K$^{+}$ ratio~\cite{gsirep98}.
Varying the repulsive potential for K$^{+}$ mesons does not modify the ratios 
significantly, as also shown in~\cite{li98}.
It has to be noted, however, that the calculations presented above   
underestimate the production yield of K$^{+}$ in the Ni+Ni collisions 
by a factor of two, while with very similar parameters 
(${\rm U}(\rho=\rho_\circ) =$+20 MeV for K$^{+}$ 
the production rate is reproduced in~\cite{chu97}. 
Despite this discrepancy, in-medium modifications of kaon masses are presently 
the only mechanism explaining the trends found in the data. 

Recently very similar observations of the 
K$^{-}$/K$^{+}$ ratios in heavy-ion collisions have been obtained at 
AGS energies (11.6 A GeV)~\cite{ogi98}. Discrepancies to transport-model
calculations were interpreted as signal for multi-body collisions owing 
to the very high densities that are reached in this energy regime.
While the sensitivity of kaon-production ratios to in-medium properties 
is enhanced in the vicinity of the elementary reaction thresholds, 
it remains to be seen by careful model analysis whether both mechanisms can 
be clearly separated. 

In summary, we measured the ratio of K$^{-}$ to K$^{+}$ meson yields as a
function of different kinematic variables in the experiments 
$^{96}$Ru+$^{96}$Ru$^{96}$/Zr at 1.69 A GeV  and 
$^{58}$Ni+$^{58}$Ni at 1.93 A GeV incident beam kinetic energy. 
We found the K$^{-}$ final-state phase-space distribution  
narrower than that of K$^{+}$ mesons.  It is unlikely that this effect 
is due to the different kinematical conditions of meson production 
or to the influence of the Coulomb potential of the net positive charge of 
colliding ions on meson propagation.  However, it can be explained when 
assuming modifications of kaon properties in a dense nuclear medium.  
Effective theories that implement and exploit chiral symmetry breaking 
patterns of QCD argue that these modifications are a consequence of the 
restoration of the symmetry in a hot and dense nuclear matter~\cite{lut94}.  
However, the drop of the K$^{-}$ effective mass can be also explained as an 
effect of the in-medium modifications of the $\Lambda(1405)$ spectral 
function due to the Pauli blocking of the proton~\cite{koc94}. 
In addition, it was suggested in \cite{schaff00} 
that in heavy ion experiments
the chaotic as well as the coherent movement of the baryonic matter
may additionally mask the influence of the in-medium potential on the
measured antikaon yields. 
In order to address the outstanding problems, 
more systematic studies of various 
systems and energies and with a better acceptance is  
needed.  A further theoretical clarifications of the origin of the 
in-medium effects on kaons is also necessary. 

We are grateful to E.L.Bratkovskaya, W.Cassing and G.Q.Li 
for providing us with the RBUU events and for useful communications. 
This work has been supported by the German BMBF under contracts 
POL-119-95, RUM-005-95, UNG-021-96 and RUS-676-98 and by the 
Deutsche Forschungsgemeinschaft
(DFG) under projects 436 RUS-113/143/2 and 446 KOR-113/76/0. 
Support has also been received from the Polish State Committee of Scientific 
Research (KBN) under grants 2P302-011-04 and 2P03B 05716, 
from the Korean Science and 
Engineering Foundation (KOSEF) under grant 985-0200-004-2, 
from the Hungarian OMFB under contract D-86/96 and
from the Hungarian OTKA under grant T029379.

\newpage

\newpage

\newpage

\vspace*{-2.cm}


\begin{thebibliography}{nel87}
\bibitem{sch97}
J.~Schaffner-Bielich, I.N.~Mishustin, J.~Bondorf, 
Nucl.\ Phys.\ {\bf A 625} (1997) 325. 

\bibitem{nel87}
A.E.~Nelson, D.B.~Kaplan, Phys.\ Lett.\ B {\bf 192} (1987) 193; \\
G.E.~Brown et al., Nucl.\ Phys.\ {\bf A 567} (1994) 937.

\bibitem{bro94}
G.E.~Brown, H.A.~Bethe, Astrophys.\ J.\ {\bf 423} (1994) 659.

\bibitem{lut94}
M.~Lutz, A.~Steiner, W.~Weise, Nucl.\ Phys.\ {\bf A 574} (1994) 755.

\bibitem{fan94}
X.S.~Fang et al., Nucl.\ Phys.\ {\bf A 575} (1994) 766.

\bibitem{sto86}
H.~St\"ocker, W.~Greiner, Phys.\ Rep.\ {\bf 137} (1986) 278.

\bibitem{bar97}
R.~Barth et al., Phys.\ Rev.\ Lett.\ {\bf 78} (1997) 4007; \\
F.~Laue et al., Phys.\ Rev.\ Lett.\ {\bf 82} (1999) 1640.

\bibitem{li95}
G.Q.~Li, C.M.~Ko, Nucl.\ Phys.\ {\bf A 594} (1995) 460; \\
E.L.~Bratkovskaya, W.~Cassing, U.~Mosel, 
Nucl.\ Phys.\ {\bf A 622} (1997) 593; \\ 
Z.S.~Wang et al., Nucl.\ Phys.\ {\bf A 628} (1998) 151.

\bibitem{rit95}
J.~Ritman et al., Z.\ Phys.\ {\bf A 352} (1995) 357; \\
P.Crochet et al., Phys.\ Lett.\ B {\bf 486} (2000) 6.

\bibitem{li97}
G.Q.~Li, C.-H.~Li, G.E.~Brown, Nucl.\ Phys.\ {\bf A 625} (1997) 372.

\bibitem{fop95}
J.~Ritman et al., Nucl.\ Phys.\ {\bf B 44} (1995) 708; \\
A.~Gobbi et al., Nucl.\ Instr.\ Meth.\ A324 (1993) 156. 

\bibitem{bes97}
D.~Best et al., Nucl.\ Phys.\ {\bf A 625} (1997) 307.

\bibitem{hon98}
B.~Hong et al., Phys.\ Rev.\  C {\bf 57} (1998) 244.

\bibitem{WIS99}
K.~Wi\'sniewski, PhD thesis, Warsaw University, (2000). 

\bibitem{brun78} 
R.~Brun et al., CERN/DD/78-2 (1978).

\bibitem{pelte97fig}
D.~Pelte et al., Z. Phys. A {\bf 359} (1997) 55.

\bibitem{ala97}
A.~Ayala, J.~Kapusta, Phys.\ Rev.\  C {\bf 56} (1997) 407; \\
H.W.~Barz et al., Phys.\ Rev.\  C {\bf 57} (1998) 2536 . 

\bibitem{li98}
G.Q.~Li, G.E.~Brown, Phys.\ Rev.\ C {\bf 58} (1998) 1698.

\bibitem{cas97}
W.~Cassing et al., Nucl.\ Phys. {\bf A 614} (1997) 415.

\bibitem{gsirep98}
K.~Wi\'{s}niewski et al., GSI Annual Report 98-1 (1998) 60.

\bibitem{chu97}
W.~Chung, G.Q.~Li, C.M.~Ko, Nucl.\ Phys. {\bf A 625} (1997) 347.

\bibitem{ogi98} 
C.A.~Ogilvie, Phys.\ Lett.\ B {\bf 436} (1998) 238. 

\bibitem{koc94}
V.~Koch, Phys.\ Lett.\ B {\bf 337} (1994) 7.

\bibitem{schaff00} 
J.~Schaffner-Bielich, V.~Koch, M.~Effenberger, 
Nucl. Phys. {\bf A 669} (2000) 153.

\end{thebibliography}
\end{document}